\documentclass[12pt]{article}

\usepackage{natbib}
\usepackage{bm}

\usepackage{amsmath}
\usepackage{amsthm}
\theoremstyle{plain}

\theoremstyle{plain}
\newtheorem{lem}{Lemma}
\theoremstyle{plain}
\newtheorem{thm}{Theorem}
\theoremstyle{remark}
\newtheorem{rem}{Remark}

\usepackage{booktabs}
\usepackage{threeparttable}
\usepackage[dvipdfmx]{graphicx}
\usepackage[hang,small,bf]{caption}
\usepackage[subrefformat=parens]{subcaption}
\usepackage{multirow}

\usepackage[dvipsnames]{xcolor}

\newcommand{\koji}[1]{\color{Bittersweet}{#1}}

\begin{document}

\def\spacingset#1{\renewcommand{\baselinestretch}%
{#1}\small\normalsize} \spacingset{1}


\title{Superset model problem}
\author{
Koji Miyawaki \\ School of Economics, Kwansei Gakuin University
\and
Steven N. MacEachern \\ Department of Statistics, The Ohio State University
}
\date{\today}
\maketitle

\begin{abstract}
This paper focuses on the superset model problem that arises in the context of regression.
To address this problem, we take the Bayesian approach to measure its uncertainty.
An illustrative example with the real dataset is provided.
%
\end{abstract}

\def\spacingset#1{\renewcommand{\baselinestretch}{#1}\small\normalsize} \spacingset{1}
\spacingset{1.5}


\section{Introduction}

Regression analysis is a statistical method widely used in many research areas.
It is often specified as the normal linear model, where coefficients are linear and the error term follows the normal distribution to simplify the analysis.
This specification aims to approximate the state of nature and is often useful in prediction as well as discussing the causality.

Among its specifications, variable selection is a central issue in the regression analysis.
It is important to select an appropriate set of explanatory variables partly because of the cost of collecting variables.
Many methods are proposed for the variable selection problem.

In relation to the variable selection problem, this paper focuses on the superset model problem where the linear regression model selects a larger set of variables than the state of nature does (see the example provided in Section \ref{sec:Superset model problem}).
The linear regression model tends to choose smaller set of variables when the state of nature is linear in variables due to its least squares loss.
However, the nonlinear relationship is more likely and this case may lead to select a larger set of variables, which will be a deficiency of the linear regression model in terms of the data collection cost.

To evaluate the superset model problem, this paper utilizes the Bayesian approach, which provides a measure of uncertainty in the form of the posterior probability, and proposes an alternative model that will select the true set of variables when the sample size is large.

This paper is organized as follows.
Section \ref{sec:Superset model problem} describes the superset model problem by providing an example.
Bayes' Theorem is adopted to evaluate the problem in Section \ref{sec:Superset model probability} and two regression models for specification is explained in Section \ref{sec:Two regression models}.
Section \ref{sec:Illustrative examples} illustrates the proposed method and discusses its robustness.


\section{Superset model problem}
\label{sec:Superset model problem}

%
Suppose the continuous response $Y$ is associated with the set of explanatory variables $\bm{x}_{T}$.
We are interested in its mean response conditional on $\bm{x}_{T}$.
We often assume it to be linear in practice, although it is more likely to be nonlinear in reality.
To this end, a regression model is specified as
\begin{align*}
Y = \phi (\bm{x}_{T}) + \epsilon,
\end{align*}
where $\epsilon$ is an additive error term with mean zero.
The functional form $\phi (\cdot)$, the distribution of the error term, and the true set of explanatory variables $\bm{x}_{T}$ are all unknown.
With this model, we make statistical inferences about the conditional mean response by estimating $\phi (\cdot)$ and the set of explanatory variables.
Among problems about how this regression model should be specified, the variable selection problem focuses on the set of explanatory variables, based on the dataset $\{ y_{i}, \bm{x}_{i} \}_{i = 1}^{n}$.

When the set of explanatory variables is known, the best fit is $E (Y \mid \bm{x}_{T})$ as an estimator of $\phi (\bm{x}_{T})$ when we use the squared loss.
The linear regression model assumes $E (Y \mid \bm{x}_{T}) = \bm{x}_{T}^{\prime} \bm{\beta}$, where $\bm{\beta}$ is referred to as the regression coefficient vector.
However, in general, $E (Y \mid \bm{x}_{T}) \neq \bm{x}_{T}^{\prime} \bm{\beta}$, contrary to the linearity assumption.

For example, suppose
\begin{align}
E \left[ Y \mid x_{T} \right] = \alpha + \beta_{1} x_{T} + \beta_{2} x_{T}^{2}.
\end{align}
The linear regression with $(x_{T}, x_{U})$, where $x_{U} = x_{T}^{2}$, is better than the one with $x_{T}$, even though the latter selects the true explanatory variable.
This is an example of the superset model problem.
On the other hand, when $x_{U}^{\prime}$ is independent of $x_{T}$, $x_{U}^{\prime}$ should not be included in the regression to improve the fit.

Above example suggests that the knowledge about association among variables is helpful to examine the superset model problem, and hence the variable selection problem.
One approach is to estimate the conditional expectation without linearity and compare it with the one implied by the normal linear model.
If they are different and the latter contains more explanatory variables, there exists the superset model problem.
Because the dataset at hand is limited, it is difficult to determine whether the superset model problem exists or not.
Rather, it is evaluated in a probability form, which is explained in the next section.


\section{Superset model probability}
\label{sec:Superset model probability}

Suppose $\mathcal{M}^{\ast}$ is the set of explanatory variables in the state of nature, which is $\bm{x}_{T}$ when vectorized, and is known for a moment.
The current dataset $\{ y_{i}, \bm{x}_{i} \}_{i = 1}^{n}$ is generated from this state of nature independently for each observation $i$ and is observed.
Depending on a context, the normal linear model with $\mathcal{M}^{\ast}$ explanatory variables would be a choice if it approximates the state of nature well.
In this case, we do not have the superset model problem.
On the other hand, a normal linear model with a set of explanatory variables indexed by $\mathcal{M} (\neq \mathcal{M}^{\ast})$ is chosen independent of the state of nature in terms of, say, prediction, where the superset model problem arises when $\mathcal{M} \supset \mathcal{M}^{\ast}$.

However, the state of nature is usually unknown and is inferred from the dataset.
The uncertainty from inference is evaluated by the posterior probability over possible subsets of explanatory variables.
This paper approximates it by assuming a flexible model (see Model \eqref{eq:alternative model} in Section \ref{sec:Two regression models}), which is denoted by $H_{0}$.
Then, this posterior probability is calculated via Bayes' theorem, which is given by
\begin{align}
\Pr \left( \mathcal{M}^{\ast} \mid \{ y_{i}, \bm{x}_{i} \}_{i = 1}^{n}, H_{0} \right)
=
\frac{ \Pr \left( \{ Y_{i} \}_{i = 1}^{n} \mid \{ \bm{x}_{i} \}_{i = 1}^{n}, H_{0}, \mathcal{M}^{\ast} \right) \Pr \left( \mathcal{M}^{\ast}, H_{0} \right) }{ \sum_{ \tilde{\mathcal{M}} } \Pr \left( \{ Y_{i} \}_{i = 1}^{n} \mid \{ \bm{x}_{i} \}_{i = 1}^{n}, H_{0}, \tilde{\mathcal{M}} \right) \Pr \left( \tilde{\mathcal{M}}, H_{0} \right) }.
\label{eq:conditional superset model probability}
\end{align}
The numerator is the cross product of the marginal likelihood and the prior belief about the set of explanatory variables.
When the latter is uniform (which is assumed in the following empirical illustration), the posterior probability is proportional to the marginal likelihood under $H_{0}$.

When uncertainty from inference about the normal linear model is evaluated from its posterior probability as well, the overall superset model probability is calculated as
\begin{align}
\sum_{ \mathcal{M} } \sum_{ \mathcal{M}^{\ast} }
I \left( \mathcal{M} \supset \mathcal{M}^{\ast} \right)
\Pr \left( \mathcal{M}^{\ast} \mid \{ y_{i}, \bm{x}_{i} \}_{i = 1}^{n}, H_{0} \right)
\Pr \left( \mathcal{M} \mid \{ y_{i}, \bm{x}_{i} \}_{i = 1}^{n}, H_{1} \right),
\label{eq:superset model probability}
\end{align}
where $H_{1}$ denotes the normal linear model (see Model \eqref{eq:normal linear regression model} in Section \ref{sec:Two regression models}).

We note that the above expression is general enough to include variables (the response and explanatory variables) that are continuous or discrete.
The next section specifies two regression models $H_{0}$ and $H_{1}$, where the response is assumed to be continuous for simplicity.


\section{Two regression models}
\label{sec:Two regression models}

First, the linear regression model $H_{1}$ is specified as
\begin{align}
Y_{i} = \alpha + \bm{x}_{i}^{\prime} \bm{\beta} + \eta_{i},
\quad
\eta_{i} \sim N \left( 0, \lambda^{2} \right),
\label{eq:normal linear regression model}
\end{align}
where each of explanatory variables is standardized without loss of generality.
To estimate model parameters $( \bm{\beta}, \lambda^{2} )$, the hyper-$g$ prior is assumed.
Then, the marginal likelihood is analytically tractable (see \citet{miyawaki-maceachern-21} for example).

Second, the model $H_{1}$ that is alternative to the normal linear regression model is specified as
\begin{align}
Y_{i} = \theta_{x} + \epsilon_{x},
\quad
\epsilon_{x} \sim N \left( 0, \sigma_{x}^{2} \right),
\label{eq:alternative model}
\end{align}
given $\bm{x}_{i} = \bm{x}$.
The normal error assumption is made because we have no other knowledge on it.
Further, it makes the conditional mean estimation simpler, in terms of the number of parameters as well as the computational burden.
The main purpose of this semiparametric model is to estimate conditional means of $Y$ in a flexible manner, and to captures the association between $Y$ and $\bm{x}$ in the state of nature as the sample size increases, which can be viewed as as an extreme of the local constant estimation (see, e.g., \citet{fan-gijbels-03} for the local estimation).

When the dataset is fixed, the covariate space becomes sparse as its dimension gets larger.
Then, the marginal likelihood (hence, the superset model probability) under this alternative model is strongly dependent on the prior specification.
To mitigate this influence, this paper takes the $m$-fold cross-validation approach, which is described in details below.

The dataset is randomly divided into $m$ groups.
One of them is used as the test set, while the remainings belong to the training set.
Explanatory variables in the training set are standardized, and those in the test set are standardized by the mean and standard deviation of those in the training set.
Let $\mathcal{D}_{0}$ and $\mathcal{D}_{1}$ be the sets of identification numbers of observations which belongs to the training and test sets.
More precisely, $\mathcal{D}_{0} = \{ i \mid \text{the $i$-th observation is in the training set}, i = 1, \dots, n \}$ and $\mathcal{D}_{1} = \{ i \mid \text{the $i$-th observation is in the test set}, i = 1, \dots, n \}$.
Given a choice of $\mathcal{D}_{0}$ and $\mathcal{D}_{1}$, we construct the prior and conditional marginal likelihood in the following manner.

The prior for $\theta_{x}$ in the model \eqref{eq:alternative model} given $\bm{x}_{i} = \bm{x}$ and $i \in \mathcal{D}_{1}$ is assumed as
\begin{align}
\theta_{x} &\sim N \left( \hat{y}_{x}, t_{x}^{2} \right),
\intertext{where $\hat{y}_{x} = \bar{y}_{0} + \bm{x}^{\prime} \hat{\bm{\beta}}$, $\bar{y}_{0}$ is the sample average of the response in $\mathcal{D}_{0}$,}
\hat{\bm{\beta}} &= \left( \sum_{i \in \mathcal{D}_{0}} \bm{x}_{i} \bm{x}_{i}^{\prime} \right)^{-1} \sum_{i \in \mathcal{D}_{0}} \bm{x}_{i} y_{i}, \\
t_{x}^{2} &= s^{2} \left\{ \frac{1}{ | \mathcal{D}_{0} | } + \bm{x}^{\prime} \left( \sum_{i \in \mathcal{D}_{0}} \bm{x}_{i} \bm{x}_{i}^{\prime} \right)^{-1} \bm{x}  \right\}, \\
\quad
s^{2} &= \frac{1}{| \mathcal{D}_{0} | - k - 1} \sum_{i \in \mathcal{D}_{0}} \left( y_{i} - \bar{y}_{0} - \bm{x}_{i}^{\prime} \hat{\bm{\beta}} \right)^{2}.
\end{align}
When $\mathcal{A}$ is a set, $| \mathcal{A} |$ is the number of elements in the set.
This prior is constructed from classical OLS estimates of mean and standard deviation of $Y_{i}$ at $\bm{x}_{i} = \bm{x}$.
By using the prior that is obtained from the linear model and using the model that focuses on the local observation, we are able to combine local and global information.
It is possible to use other estimates such as corresponding normal linear regression estimates under the hyper-g prior.
However, to keep the methodology as simple as possible, we take the above prior specification.

Then, we are able to derive the marginal likelihood conditional on the nuisance parameter $\sigma_{x}^{2}$ for each $\bm{x}_{i} = \bm{x}$ and $i \in \mathcal{D}_{1}$.
Let $\mathcal{M}_{x} = \{ i \mid \bm{x}_{i} = \bm{x}, i \in \mathcal{D}_{1} \}$ and $n_{x} = | \mathcal{M}_{x} |$.
Then, this conditional marginal likelihood is given by
\begin{align*}
&m^{\ast} \left( \{ Y_{i} \}_{i \in \mathcal{M}_{x} } \mid \{ \bm{x}_{i} \}_{i \in \mathcal{M}_{x} }, \sigma_{x}^{2}, \{ y_{i}, \bm{x}_{i} \}_{i \in \mathcal{D}_{0}} \right) \notag \\
&\hspace{100pt} =
\frac{ \tau_{x} }{ \left( \sqrt{2 \pi} \sigma_{x} \right)^{n_{x}} t_{x} }
\exp \left\{ -\frac{1}{2} \left( -\frac{\mu_{x}^{2}}{\tau_{x}^{2}} + \frac{ \sum_{i \in \mathcal{M}_{x}} y_{i}^{2} }{\sigma_{x}^{2}} + \frac{ \hat{y}_{x}^{2} }{ t_{x}^{2} } \right) \right\}, \\
&\mu_{x} = \tau_{x}^{2} \left( \frac{ \sum_{i \in \mathcal{M}_{x}} y_{i} }{\sigma_{x}^{2}} + \frac{\hat{y}_{x}}{t_{x}^{2}} \right),
\quad
\tau_{x}^{2} = \left( \frac{n_{x}}{\sigma_{x}^{2}} + \frac{1}{t_{x}^{2}} \right)^{-1}.
\end{align*}

The full Bayes analysis specifies a prior on the nuisance parameter $\sigma_{x}^{2}$ as well.
However, because the data are sparse at $\bm{x}$, how we specify it affects the (unconditional) marginal likelihood much.
To mitigate this problem, this paper takes the empirical Bayes approach.
The marginal likelihood for each $\bm{x}_{i} = \bm{x}$ is the conditional marginal likelihood $m^{\ast} ( \{ Y_{i} \}_{i \in \mathcal{M}_{x} } \mid \{ \bm{x}_{i} \}_{i \in \mathcal{M}_{x} }, \sigma_{x}^{2}, \{ y_{i}, \bm{x}_{i} \}_{i \in \mathcal{D}_{0}} )$ maximized over $\sigma_{x}^{2}$.
More precisely,
\begin{align*}
m \left( \{ Y_{i} \}_{i \in \mathcal{M}_{x} } \mid \{ \bm{x}_{i} \}_{i \in \mathcal{M}_{x} }, \{ y_{i}, \bm{x}_{i} \}_{i \in \mathcal{D}_{0}} \right)
\equiv
\max_{\sigma_{x}^{2}}
m^{\ast} \left( \{ Y_{i} \}_{i \in \mathcal{M}_{x} } \mid \{ \bm{x}_{i} \}_{i \in \mathcal{M}_{x} }, \sigma_{x}^{2}, \{ y_{i}, \bm{x}_{i} \}_{i \in \mathcal{D}_{0}} \right).
\end{align*}
See the next section for this maximization in details.
By multiplying it over all distinct $\bm{x}$ in $\mathcal{D}_{1}$ and taking the geometric mean, we have the marginal likelihood for $\mathcal{D}_{1}$ per one observation, which is given by
\begin{align*}
\left\{
\prod_{\bm{x}}
m \left( \{ Y_{i} \}_{i \in \mathcal{M}_{x} } \mid \{ \bm{x}_{i} \}_{i \in \mathcal{M}_{x} }, \{ y_{i}, \bm{x}_{i} \}_{i \in \mathcal{D}_{0}} \right)
\right\}^{1 / | \mathcal{D}_{1} |}.
\end{align*}
The geometric mean is to take care of different sample sizes in different test sets.

Finally, we repeat above process until all $m$ groups are used as the test set and calculate above marginal likelihood for each test group selection.
After averaging $m$ marginal likelihoods, we raise it to the power of $n$ to obtain the final marginal likelihood estimate for the model \eqref{eq:alternative model}.

The robustness of this approach is of interest.
The approach will be more useful if we know the upper and lower bounds of the superset model probability under $H_{0}$ when its specification changes.
Two points are discussed regarding robustness.

\begin{figure}[ht]
\centering
\includegraphics[width=12cm, clip, keepaspectratio]{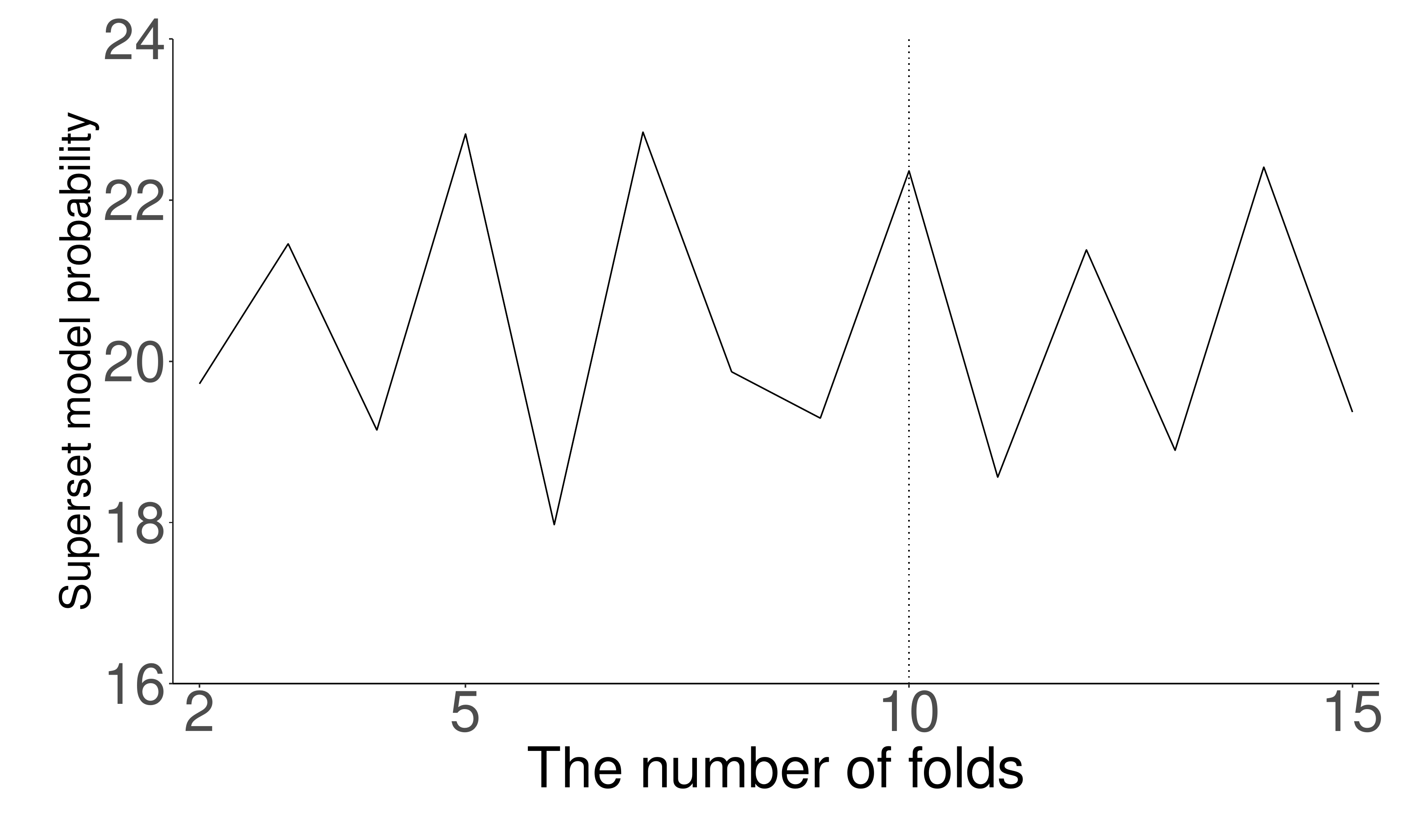}
\caption{Superset model probability change as the number of folds increases. (The dotted line indicates the 10-fold cross-validation, which will be used in Section \ref{sec:Illustrative examples})}
\label{fig:Superset model probability change as the number of folds increases.}
\end{figure}

First, we consider the robustness to the number of folds in the cross-validation.
In the following empirical analysis in Section \ref{sec:Illustrative examples}, we use the 10-fold cross-validation.
When the number of folds changes from 2 to 15, we see the resulting probability does not change much with the diabetes dataset (see Figure \ref{fig:Superset model probability change as the number of folds increases.}).

Second, it is important to check the robustness to the prior.
One approach would be to use the $\epsilon$-contamination class prior (see Section 4.7.4 of \citet{berger-85}), and to show the sensitivity of the marginal likelihood, which will be our future analysis.


\section{Maximize the marginal likelihood}
\label{sec:Maximize the marginal likelihood}


%
%
Letting
\begin{align}
\bar{y}_{x} &= \frac{1}{n_{x}} \sum_{i \in G_{x}} y_{i}, \\
s_{y}^{2} &= \frac{1}{n_{x}} \sum_{i \in G_{x}} \left( y_{i} - \bar{y}_{x} \right)^{2}
           = \frac{1}{n_{x}} \sum_{i \in G_{x}} y_{i}^{2} - \bar{y}_{x}^{2},
\end{align}
the local marginal likelihood function is characterized by Theorem \ref{thm:maximization}.
\begin{thm}
The local marginal likelihood function has extremal values at
positive solutions to the cubic equation \eqref{eq:cubic equation} given in the proof when $s_{y}^{2} > 0$;
at 0 when $s_{y}^{2} = 0, t_{x}^{2} - ( \bar{y}_{x} - \hat{y}_{x} )^{2} \geq 0$; and
at 0 and $- t_{x}^{2} + ( \bar{y}_{x} - \hat{y}_{x} )^{2}$ when $s_{y}^{2} = 0, t_{x}^{2} - ( \bar{y}_{x} - \hat{y}_{x} )^{2} < 0$.
\label{thm:maximization}
\end{thm}
See Appendix \ref{sec:Proof of Theorem} for its proof.
Table \ref{table:Local marginal likelihood function} summarizes the result.
\begin{table}
\centering
\caption{Local marginal likelihood function}
\label{table:Local marginal likelihood function}
\begin{tabular}{lccc}
\toprule
                & $\sigma_{x}^{2} \to 0$ & $0 < \sigma_{x}^{2} < \infty$ & $\sigma_{x}^{2} \to \infty$ \\
\midrule
$s_{y}^{2} > 0$ & 0 & \multirow{2}{*}{See Lemma \ref{lem:extremal values}} & \multirow{2}{*}{0} \\
$s_{y}^{2} = 0$ & $\frac{1}{ \sqrt{2 \pi} t_{x}} \exp \left\{ -\frac{1}{2 t_{x}^{2}} \left( \bar{y}_{x} - \hat{y}_{x} \right)^{2} \right\}$ & & \\
\bottomrule
\end{tabular}
\end{table}
By this theorem, we are able to set a value of $\sigma_{x}^{2}$ to maximize the marginal likelihood, instead of placing a prior on it.


\section{Illustrative example}
\label{sec:Illustrative examples}

The diabetes data (see \citet{efron-etal-04}) are used to illustrate our method.
This dataset contains 442 observations.
For the analysis below, we use the logarithm of the diabetes progression measure as the response and use remaining 10 variables are used as exlanatory variables.
The proposed method is applied, and the superset model probability for this dataset with 10-fold cross-validation is estimated to be 22.37\%.

As discussed by \citet{maceachern-miyawaki-22}, the diabetes dataset seems to be collected from least two different sources.
In particular, the precision of two explanatory variables (the blood pressure and fourth blood serum measurement) consists of a mix of finer and coarser observations.
When the data are divided into two groups by this precision, \citet{maceachern-miyawaki-22} suggests these two datasets have different characteristics.

This conclusion is also confirmed in terms of the superset model probability.
When the dataset for observations with finer variables is used, it is 22.19\%.
When, on the other hand, that for observations with coarser variables is used, it is 10.72\%.
The superset model problem is more likely to occur with the former dataset than the latter one, which would be due to the difference in characteristics of these two datasets.






\appendix
\section{Proof of Theorem \ref{thm:maximization}}
\label{sec:Proof of Theorem}

The local marginal likelihood function is characterized in the following two lemmas.
The proof of the theorem directly follows from them.

\begin{lem}
The local marginal likelihood function converges to a finite value as $\sigma_{x}^{2}$ approaches zero or infinity.
\end{lem}
\begin{proof}
Observe that
\begin{align}
g \left( \sigma_{x}^{2} \right)
&=
-\frac{\mu_{x}^{2}}{\tau_{x}^{2}} + \frac{ \sum_{i \in G_{x}} y_{i}^{2} }{\sigma_{x}^{2}} \notag \\
&=
\frac{1}{\sigma_{x}^{2}} \left\{ \sum_{i \in G_{x}} y_{i}^{2} - \frac{\tau_{x}^{2}}{\sigma_{x}^{2}} \left( \sum_{i \in G_{x}} y_{i} \right)^{2} \right\}
- 2 \frac{\tau_{x}^{2}}{\sigma_{x}^{2}} \frac{n_{x} \bar{y}_{x} \hat{y}_{x}}{t_{x}^{2}}
- \tau_{x}^{2} \frac{\hat{y}_{x}^{2}}{t_{x}^{4}} \notag \\
&=
\frac{1}{\sigma_{x}^{2}} \left\{ n_{x} s_{y}^{2} + \left( \frac{1}{n_{x}} - \frac{\tau_{x}^{2}}{\sigma_{x}^{2}} \right) n_{x}^{2} \bar{y}_{x}^{2}
\right\}
- 2 \frac{\tau_{x}^{2}}{\sigma_{x}^{2}} \frac{n_{x} \bar{y}_{x} \hat{y}_{x}}{t_{x}^{2}}
- \tau_{x}^{2} \frac{\hat{y}_{x}^{2}}{t_{x}^{4}}.
%
\end{align}

First, we consider the convergence when $\sigma_{x}^{2} \to \infty$.
Because $\tau_{x}^{2} \to t_{x}^{2}$ and $\frac{\tau_{x}^{2}}{\sigma_{x}^{2}} \to 0$, $g (\sigma_{x}^{2}) \to -\frac{ \hat{y}_{x}^{2} }{ t_{x}^{2} }$, as $\sigma_{x}^{2} \to \infty$.
Hence, the local marginal likelihood function converges to zero in this case.

Next, we consider the case when $\sigma_{x}^{2} \to 0$.
As $\sigma_{x}^{2} \to 0$, $\tau_{x}^{2} \to 0$ and $\frac{\tau_{x}^{2}}{\sigma_{x}^{2}} \to \frac{1}{n_{x}}$.
Because
\begin{align}
\frac{1}{\sigma_{x}^{2}} \left( \frac{1}{n_{x}} - \frac{\tau_{x}^{2}}{\sigma_{x}^{2}} \right)
&=
\frac{1}{n_{x} \left( n_{x} t_{x}^{2} + \sigma_{x}^{2} \right) },
\intertext{we have}
g \left( \sigma_{x}^{2} \right) &\to
\begin{cases}
\infty, &\text{when $s_{y}^{2} > 0$}, \\
\frac{ \bar{y}_{x}^{2} }{ t_{x}^{2}} - 2 \frac{ \bar{y}_{x} \hat{y}_{x} }{ t_{x}^{2} }, &\text{when $s_{y}^{2} = 0$},
\end{cases}
\end{align}
as $\sigma_{x}^{2} \to 0$.
Thus, the local marginal likelihood function converges to
\begin{align}
\begin{cases}
0, &\text{when $s_{y}^{2} > 0$}, \\
\frac{1}{ \sqrt{2 \pi} t_{x}} \exp \left\{ -\frac{1}{2 t_{x}^{2}} \left( \bar{y}_{x} - \hat{y}_{x} \right)^{2} \right\}, &\text{when $s_{y}^{2} = 0$},
\end{cases}
\end{align}
when $\sigma_{x}^{2} \to 0$.
\end{proof}

\begin{lem}
For $0 < \sigma_{x}^{2} < \infty$, at least one positive extremal value of the local marginal likelihood function exists when $s_{y}^{2} > 0$, while at most one such a value exists otherwise.
\label{lem:extremal values}
\end{lem}
\begin{proof}
The first order condition for maximizing the log local marginal likelihood function is
\begin{align}
\frac{
m^{\prime} \left( \{ Y_{i} \}_{i \in G_{x} } \mid \{ \bm{x}_{i} \}_{i \in G_{x} }, \sigma_{x}^{2} \right)
}
{
m \left( \{ Y_{i} \}_{i \in G_{x} } \mid \{ \bm{x}_{i} \}_{i \in G_{x} }, \sigma_{x}^{2} \right)
}
=
0.
\end{align}
The left hand side is calculated as
\begin{align}
&\frac{1}{2} \frac{ \left( \tau_{x}^{2} \right)^{2} }{ \left( \sigma_{x}^{2} \right)^{4} } n_{x}
\left\{ a_{1} \left( \sigma_{x}^{2} \right)^{3} + a_{2} \left( \sigma_{x}^{2} \right)^{2} + a_{3} \sigma_{x}^{2} + a_{4} \right\},
\intertext{where}
&a_{1} = -\frac{1}{t_{x}^{4}},
\quad
a_{2} = \frac{1}{t_{x}^{2}} \left( 1 - 2 n_{x} \right) + \frac{1}{t_{x}^{4}} \left\{ s_{y}^{2} + \left( \bar{y}_{x} - \hat{y}_{x} \right)^{2} \right\}, \\
&a_{3} = -n_{x}^{2} + n_{x} + \frac{2 n_{x} s_{y}^{2}}{t_{x}^{2}},
\quad
a_{4} = n_{x}^{2} s_{y}^{2},
\end{align}
Then, the extremal values are solutions to the following cubic equation:
\begin{align}
a_{1} \left( \sigma_{x}^{2} \right)^{3} + a_{2} \left( \sigma_{x}^{2} \right)^{2} + a_{3} \sigma_{x}^{2} + a_{4} = 0.
\label{eq:cubic equation}
\end{align}

When $s_{y}^{2} > 0$, $a_{4} > 0$.
Thus, at least one solution to this cubic equation is positive when $s_{y}^{2} > 0$.
\begin{rem}
Let $(\alpha, \beta, \gamma)$ be three solutions to this cubic equation.
Then, it is factorized as $(\sigma_{x}^{2} - \alpha) (\sigma_{x}^{2} - \beta) (\sigma_{x}^{2} - \gamma) = 0$.
By comparing coefficients, $-\alpha \beta \gamma = \frac{a_{4}}{a_{1}} < 0$.
This implies that solutions are nonzero and either one of three possibilities: (i) three distinct real solutions, (ii) one real and two complex conjugate solutions, or (iii) three real, but at least two are the same, solutions.
Then, we conclude that at least one solution is positive.
\end{rem}

When $s_{y}^{2} = 0$, the cubic equation reduces to
\begin{align}
\left( \sigma_{x}^{2} \right)^{3} + \left\{ t_{x}^{2} - \left( \bar{y}_{x} - \hat{y}_{x} \right)^{2} \right\} \left( \sigma_{x}^{2} \right)^{2} = 0.
%
\end{align}
If $t_{x}^{2} - ( \bar{y}_{x} - \hat{y}_{x} )^{2} \geq 0$, no extremal value exists for the local marginal likelihood function over $0 < \sigma_{x}^{2} < \infty$.
If, on the other hand, $t_{x}^{2} - ( \bar{y}_{x} - \hat{y}_{x} )^{2} < 0$,
\begin{align}
\sigma_{x}^{2} = - t_{x}^{2} + \left( \bar{y}_{x} - \hat{y}_{x} \right)^{2}
\end{align}
is the extremal value within the same range.
\end{proof}


\bibliographystyle{chicago}
\bibliography{superset_bib}


\end{document}